\documentclass[12pt]{amsart}

\usepackage{amsaddr}
\usepackage[authoryear,round]{natbib}
\usepackage{float}
\usepackage{hyperref}
\usepackage[pdftex]{graphicx}
\usepackage{algorithm}
\usepackage{alltt}
\usepackage{caption}
\usepackage[a4paper, margin=2cm]{geometry}

\newtheorem{thm}{Theorem}

\theoremstyle{remark}\newtheorem*{rmk}{Remark}
\newtheorem*{pot1}{Proof of Theorem \ref{T1}}
\newtheorem*{pot2}{Proof of Theorem \ref{T2}}
\newtheorem*{pot3}{Proof of Theorem \ref{T3}}

\begin{document}

\title{A Note on Cohen's d From a Partitioned Linear Regression Model}

\author{J\"{u}rgen Gro{\ss}}
\address{Institute for Mathematics and Applied Informatics, University of Hildesheim, Germany}
\email{juergen.gross@uni-hildesheim.de}
\author{Annette M\"{o}ller}
\address{Faculty of Business Administration and Economics,  Bielefeld University, Germany}
\email{annette.moeller@uni-bielefeld.de}
\thanks{Support of the second author by the Helmholtz Association’s pilot project ”Uncertainty Quantification” is gratefully acknowledged.}



\date{}

\begin{abstract} In this note we introduce a generalized formula for Cohen's $d$ under the presence of additional independent variables, providing a measure for the size of a possible effect concerning the location difference of a variable in two groups. This is done by employing the so-called Frisch-Waugh-Lovell theorem in a partitioned linear regression model. The generalization is motivated by demonstrating  the relationship to appropriate $t$ and $F$ statistics. Our discussion is further illustrated by inference from a publicly available data set.
\end{abstract}

\maketitle

\markboth{J.~Gro{\ss} \& A. M\"{o}ller}{Cohen's d From a Partitioned Linear Regression Model}

\section{Introduction}\label{sec:intro}

When applying statistical testing of hypotheses to data it is often recommended not only to report the corresponding p-value, but in addition to provide a measure for the effect associated with a possible rejection of the null hypothesis, see e.g. \citet{wilkinson1999statistical}. Such a measure may be useful when sample sizes are to be fixed during the planning phase of a study, or when it is desired to assess the  relevance of an actual rejection when given sample sizes are large. Effect size measures are strongly related to power analysis as carried out in the seminal book by \citet{cohen1988statistical}.

A widely used measure is the so-called Cohen's $d$, see also \citet{hedges1981distribution, kraemer1983theory}, which is an effect size measure for the two-sample $t$ test with equal variances. Consider independent samples of sizes $n_{1}$ and $n_{2}$ of a statistical variable $y$ in two groups such that
$y$ follows a normal distribution with expectation $\mu_{1}$ and variance $\sigma^2$ in group 1 and expectation $\mu_{2}$ and the same variance $\sigma^2$ in group 2. Let $t$ denote the usual two-sample test statistic for the null hypotheses
$H_{0}: \mu_{1} = \mu_{2}$ versus the alternative $H_{1}: \mu_{1} \not = \mu_{2}$. As a measure for the size
of an effect, \citet[p. 66ff]{cohen1988statistical} considers the absolute value of
\begin{equation}\label{E1}
d = \frac{\overline{y}_{1}-\overline{y}_{2}}{\sqrt{\frac{s_{1}^2 + s_{2}^2}{n_{1} + n_{2} -2}}}\; ,
\end{equation}
where $\overline{y}_{j}$ is the sample mean in group $j$, $j=1,2$, and $s_{j}^2 =\sum_{i} (y_{i} - \overline{y}_{j})^2$, where summation is carried out with respect to all observation from group $j$. The effect size $d$ is related to the test statistic $t$ by the formula
\begin{equation}\label{E2}
d = t \sqrt{\frac{n_{1}+ n_{2}}{n_{1} n_{2}}}\; ,
\end{equation}
see (2.5.3) in \citet{cohen1988statistical}.
According to \citeauthor{cohen1988statistical}, values $|d|=0.2$, $|d|=0.5$ and $|d|=0.8$ indicate a small, medium and large effect, respectively.

It may also be of interest to have a corresponding measure when the variable $y$ depends on further independent variables. In his Chapter 9, \citet{cohen1988statistical} deals with such a multiple regression situation and discusses the effect size measure $f^2$ at length, as will further be explicated in our Section \ref{C1}.

However, an analogous measure to $d$ is rare to find, see
\citet[Sect. 3.14]{wilson2016formulas}, \citet{lipsey2001practical} for such a proposal. Nonetheless, it may be of particular interest to have comparable measures of an effect size for the very same  grouped variable $y$  but additionally depending on different sets of independent variables. This is exemplarily carried out in our Section \ref{C2}.
In the following we introduce such a measure as a generalization to $d$ by considering a linear regression model
\begin{equation}\label{E3}
y = \beta_{0}+ \beta_{1} z + \beta_{2} x_{1} + \cdots + \beta_{w+1} x_{w} +\varepsilon\; ,
\end{equation}
where $z$ takes the value $z_{i} = 0$ if the corresponding observation $y_{i}$ of the dependent variable $y$ belongs to group 1 and $z_{i} = 1$ if $y_{i}$ belongs to group 2, $i= 1,\ldots, n_{1} + n_{2}$. It is assumed that there are $w$ independent variables $x_{1}, \ldots, x_{w}$. %, which in case $w=0$ are considered as absent%, thus referring to the simple model $y = \beta_{0}+ \beta_{1} z  +\varepsilon$.
The error variable $\varepsilon$ is assumed to follow a normal distribution with expectation $0$ and variance $\sigma^2$.

As will be shown in the following Sections \ref{sec:part}, \ref{C3}, and \ref{C1}, a natural generalization of Cohen's $d$ is given by
\begin{equation}\label{E9}
d_{\ast} = \frac{\overline{y_{\ast}}_{1}-\overline{y_{\ast}}_{2}}{\sqrt{\frac{s_{\ast 1}^2 + s_{\ast 2}^2}{n_{1} + n_{2} - 2 - w}}}, \quad  s_{\ast j}^2 = \sum_{i} (y_{\ast i} - \overline{y_{\ast}}_{j})^2, \quad j=1,2\; ,
\end{equation}
where $y_{\ast} = y - \widehat{\beta}_{2} x_{1} - \cdots - \widehat{\beta}_{w+1} x_{w}$ is the dependent variable adjusted for the independent variables. The $\widehat{\beta}_{k}$ are the ordinary least squares estimates  of the regression coefficients $\beta_{k}$, $k=2,\ldots, w+1$ in model (\ref{E3}). In case $w=0$, the adjusted $y_{\ast}$ coincides with the original $y$, so that (\ref{E9}) reduces to (\ref{E1}) and therefore can be seen as a natural generalization of Cohen's $d$.

\section{Partitioned Linear Regression}\label{sec:part}

Let $n= n_{1} + n_{2}$ be the total sample size.  The above model (\ref{E3}) may also be written in vector-matrix notation as
\begin{equation}\label{E4}
y = X_{1} \delta_{1} + X_{2} \delta_{2} + \varepsilon, \quad
\end{equation}
where now $y$ represents the $n \times 1$ vector of observations of the dependent variable. Without loss
of generality it is assumed that the first $n_{1}$ observations belong to group 1, while the last $n_{2}$ observations belong to group 2. By introducing the notation $1_{m}$ for an $m\times 1$ vectors of ones, the $n\times 2$ matrix $X_{1}$ and the corresponding  $2\times 1$ parameter vector $\delta_{1}$ may be written as
\begin{equation}\label{E5}
X_{1} = \begin{pmatrix}
1_{n_{1}} & 0\\
1_{n_{2}} & 1_{n_{2}}
\end{pmatrix} \quad \text{and}\quad
\delta_{1} = \begin{pmatrix}
\beta_{0}\\
\beta_{1}
\end{pmatrix}\; .
\end{equation}
The $n\times w$ matrix $X_{2}$ contains the observations of the independent variables with corresponding regression coefficients $\delta_{2}^{T} = (\beta_{2}, \ldots, \beta_{w+1})$, where the $T$ superscript denotes transposition. The $n\times 1$ random vector $\varepsilon$ is assumed to follow a  multivariate normal distribution with expectation vector $0$ and variance-covariance matrix $\sigma^2 I_{n}$, where $I_{n}$ stands for
the $n\times n$ identity matrix. It is assumed that the $n\times (2+w)$ model matrix $(X_{1}, X_{2})$ has full column rank
$2 +w$. Equation (\ref{E4}) represents a partitioned linear regression model as considered e.g. in
\citet{fiebig1996frisch}. Generalizations and further properties are investigated  by
\citet{puntanen1996some, gross2000estimation, gross2005extensions, ding2021frisch}, among others.

Under model (\ref{E4}) the ordinary least squares estimator for the parameter vector $(\delta_{1}^{T}, \delta_{2}^{T})$ is given by
\begin{equation}\label{E6}
\begin{pmatrix}
\widehat{\delta}_{1}\\
\widehat{\delta}_{2}
\end{pmatrix}  = (X^{T} X)^{-1} X^{T} y,  \quad X=(X_{1},  X_{2})\; .
\end{equation}
The Frisch-Waugh-Lovell theorem, see \citet{fiebig1996frisch,lovell1963seasonal,frisch1933partial}, states that
\begin{equation}\label{E7}
\widehat{\delta}_{2} = (X_{2}^{T} M_{1}  X_{2})^{-1} X_{2}^{T} M_{1} y,\quad M_{1} = I_{n}- X_{1} (X_{1}^{T} X_{1})^{-1} X_{1}^{T}\; .
\end{equation}
For the specific choice (\ref{E5}), the matrix $M_{1}$ becomes
\begin{equation}\label{E8}
M_{1} = \begin{pmatrix}
C_{1} & 0\\
0 & C_{2}
\end{pmatrix},\quad C_{j} = I_{n_{j}} - n_{j}^{-1} 1_{n_{j}} 1_{n_{j}}^{T},\quad j=1,2\; .
\end{equation}

The following result is not  restricted to the case (\ref{E5}) but remains valid
in situations where  the matrix $X_{1}$ corresponds to an arbitrary set of $v$ independent variables such that the assumptions of (\ref{E4}) are satisfied.

\begin{thm}\label{T1} Under the partitioned linear regression model (\ref{E4}),
\begin{equation}
\widehat{\delta}_{1} = (X_{1}^{T}X_{1})^{-1} X_{1}^{T} (y - X_{2} \widehat{\delta}_{2})
\end{equation}
is the ordinary least squares estimator of $\delta_{1}$.
\end{thm}

A proof is given in the appendix. Theorem \ref{T1} means that if $\widehat{\delta}_{2}$ is known (e.g. computed by
(\ref{E7})), then the remaining parameters $\delta_{1}$ can  be estimated by regressing the adjusted \begin{equation}
y_{\ast} = y - X_{2} \widehat{\delta}_{2}\end{equation}  on the remaining  $X_{1}$ and this procedure just yields the identical estimate of $\delta_{1}$ from (\ref{E6}).

\begin{thm}\label{T2} Under the partitioned linear regression model (\ref{E4}) and (\ref{E5}),
\begin{eqnarray}
\widehat{\sigma}^2 & = & (y - X_{2} \widehat{\delta}_{2})^{T} M_{1} (y - X_{2} \widehat{\delta})/(n-2-w)\label{T2E1}\\
& = & (s_{\ast1}^{2} + s_{\ast 2}^{2})/(n-2-w)\label{T2E2}
\end{eqnarray}
is an unbiased estimator for $\sigma^2$.
\end{thm}

As a matter of fact, $\widehat{\sigma}^2$ coincides with the usual estimator for $\sigma^2$ in model (\ref{E4}). Identity (\ref{T2E2}) follows immediately from (\ref{E8}) with the above definition of $y_{\ast}$.

\section{Testing for a group effect}\label{C3}

From Theorem \ref{T1} with  $X_{1}$ from (\ref{E5}) it follows that
\begin{equation}\label{E18}
\widehat{\delta}_{1} =  \begin{pmatrix}
\widehat{\beta}_{0}\\
\widehat{\beta}_{1}
\end{pmatrix}= \begin{pmatrix}
n_{1}^{-1} 1_{n_{1}} & 0\\
- n_{1}^{-1} 1_{n_{1}} & n_{2}^{-1} 1_{n_{2}}
\end{pmatrix} y_{\ast}
= \begin{pmatrix}
\overline{y_{\ast}}_{1}\\
\overline{y_{\ast}}_{2} - \overline{y_{\ast}}_{1}
\end{pmatrix}\; .
\end{equation}
Hence, it is seen that $|d_{\ast}|$ from (\ref{E9}) is identical to
\begin{equation}\label{E17}
|d_{\ast}| = \frac{|\widehat{\beta}_{1}|}{\widehat{\sigma}}
\end{equation}
with $\widehat{\sigma}$ being the square root of $\widehat{\sigma}^2$ from Theorem \ref{T2}. The
statistic $d_{\ast}$ is closely related to the test statistic $t_{\ast}$ for the null hypothesis
$H_{0}: \beta_{1} = 0$ in model (\ref{E4}).

\begin{thm}\label{T3} Under the partitioned linear regression model (\ref{E4}) and (\ref{E5}) let
$M_{2} = I_{n} - P_{2}$, $P_{2} = X_{2} (X_{2}^{T}  X_{2})^{-1} X_{2}^{T}$, and let
$\gamma$ be the lower-right element of the $2\times 2$ matrix $(X_{1}^{T} M_{2} X_{1})^{-1}$. Then the  statistic
\begin{equation}\label{E10}
t_{\ast} = d_{\ast} /\sqrt{\gamma}
\end{equation}
follows a central $t$ distribution with $n - 2-w$ degrees of freedom, provided $\beta_{1} =0$.
\end{thm}

In the above theorem, $\gamma$ is the scaled variance of $\widehat{\beta}_{1}$, i.e. $\text{Var}(\widehat{\beta}_{1}) = \sigma^2 \gamma$, see the proof of Theorem \ref{T3} in the appendix. The standard error of
$\widehat{\beta}_{1}$ is thus $\text{se}(\widehat{\beta}_{1}) = \widehat{\sigma} \sqrt{\gamma}$ with $\widehat{\sigma}$ being the square root of $\widehat{\sigma}^2$ from Theorem \ref{T2}.

Note that in case $w = 0$ by setting $M_{2}=I_{n}$ one gets
\begin{equation}
(X_{1}^{T} X_{1})^{-1} = \begin{pmatrix}
n_{1}^{-1} & - n_{1}^{-1}\\
- n_{1}^{-1} & \gamma
\end{pmatrix}, \quad \gamma
= \frac{n_{1} + n_{2}}{n_{1} n_{2}}\; ,
\end{equation}
and hence
\begin{equation}
d_{\ast} = t_{\ast} \sqrt{\frac{n_{1} + n_{2}}{n_{1} n_{2}}}\; ,
\end{equation}
which is just a reformulation of (\ref{E2}). These considerations show that $d_{\ast}$ is a natural extension
of Cohen's $d$ in the context of additional independent variables.

\section{Effect Size in Multiple Regression}\label{C1}

In his Chapter 9, \citet{cohen1988statistical} discusses the effect size measure $f^{2}$ based on the $F$ test of a linear hypothesis. It may be applied  when $X_{1}$ does not only comprise intercept and one dummy as under model (\ref{E4}), but a total of $u$ independent variables. Then it might be of interest to measure the effect size of the set of variables in $X_{1}$ given the set in $X_{2}$, which is Cohen's case 1. \citeauthor{cohen1988statistical} suggests values $f^2=0.02$, $f^2=0.15$ and $f^2=0.35$ for a small, medium and large effect, respectively.
Since the measure $d_{\ast}$ refers to one dummy ($u=1$), one might expect a relationship between $d_{\ast}$ and the corresponding $f^2$.  Actually, as noted in our Remark below, such a relationship can be specified.

The measure $f^2$ for Cohen's case 1 is  given by
\begin{equation}\label{E11}
f^2  = F \frac{u}{v},\quad  v = n-u-w- 1\; ,
\end{equation}
where under model (\ref{E4}) $F$ is the $F$ statistic for testing the null hypothesis $H_{0}: \beta_{1}=0$. From (9.2.3) in \citet{cohen1988statistical},
\begin{equation}\label{E15}
f^2 = \frac{R^2 - R_{0}^2}{1 -R^2}\; ,
\end{equation}
where $R^2$ is the coefficient of determination from model (\ref{E4}) and $R_{0}^2$ is the coefficient of determination in the reduced model with $\beta_{1}=0$, admitting model matrix $X_{3}=(1_{n}, X_{2})$. If $P$ denotes
the orthogonal projector onto the column space of the model matrix of a regression model with intercept, the coefficient of determination is given by
\begin{equation}
R^2 = 1 - \frac{y^{T} (I_{n} - P) y}{y^{T} C y}%, \quad R_{0}^2 = 1 - \frac{y^{T} (I_{n} - P_{3}) y}{y^{T} C y}
\end{equation}
with $C= I_{n} - n^{-1} 1_{n} 1_{n}^{T}$ being the so-called centering matrix, e.g. see \citet[Sect. 6.2]{gross2003linear}. From this, (\ref{E15}) becomes
\begin{equation}
f^2 = \frac{y^{T}(P - P_{3}) y}{y^{T} (I_{n} - P) y}
\end{equation}
with $P=X(X^{T} X)^{-1} X^{T}$, $X=(X_{1}, X_{2})$, and $P_{3} = X_{3} (X_{3}^{T} X_{3})^{-1} X_{3}^{T}$.
In view of $\text{rank}(P-P_{3}) =1$ and $\text{rank}(I_{n} - P) = n - (2 + w)$, the corresponding $F$ statistic reads
\begin{equation}
F = \frac{y^{T}(P - P_{3}) y/\text{rank}(P-P_{3})}{y^{T} (I_{n} - P) y/\text{rank}(I_{n} - P)}\; .
\end{equation}
Then, from Theorem 3.2.1 (ii) in \citet{christensen2020plane}, $F$ follows a central $F$ distribution with  $1$ and $n-2-w$ degrees of freedom, provided $\beta_{1}=0$.

Now, it is well known and readily verified that the squared $t$ statistic for the null hypothesis $H_{0}: \beta_{1}=0$ is identical to the test statistic of the $F$ test for the very same hypothesis. Thus, by combining
(\ref{E10}) and (\ref{E11}) the following is true.

\begin{rmk} The identity
\begin{equation}\label{E14}
f^2 = \frac{d_{\ast}^2 /\gamma}{n-2-w}
\end{equation}
specifies the exact relationship between the effect size measures $f^2$ and $d_{\ast}$ from above.
\end{rmk}

Note that in case $w=0$ the above identity reads
\begin{equation}\label{E12}
f^2 = d^{2} \frac{n_{1} n_{2}}{n (n-2)}\; ,
\end{equation}
which slightly differs from formula (9.3.5) in \citet{cohen1988statistical} and lacks some of its beauty. Since (\ref{E12})
is expected to coincide with (9.3.5) this reveals a fallacy in the latter formula. Formula (\ref{E12}) may also be verified independently by directly assuming model (\ref{E4}) without any additional independent variables $X_{2}$. In most cases actual computations of the two formulas in question only differ in a later digit after the dot (say the fifth or sixth), so the difference usually has no practical meaning. The correctness of (\ref{E14}) and (\ref{E12}) is additionally confirmed by  applications to real data.

\section{Data Example}\label{C2}

To give a possible outline for applications and an illustration of the previous formulas we employ a data set available from the UCI machine learning repository, see \citet{Dua:2019}. It contains
student achievement in secondary education of two Portuguese schools, see \citet{cortez2008using}. In the following, computations are carried out with the  statistical software {\sf R} \citep{Rsoftware}.

\begin{figure}[hbt]
\centering
\includegraphics[width=19pc,angle=0]{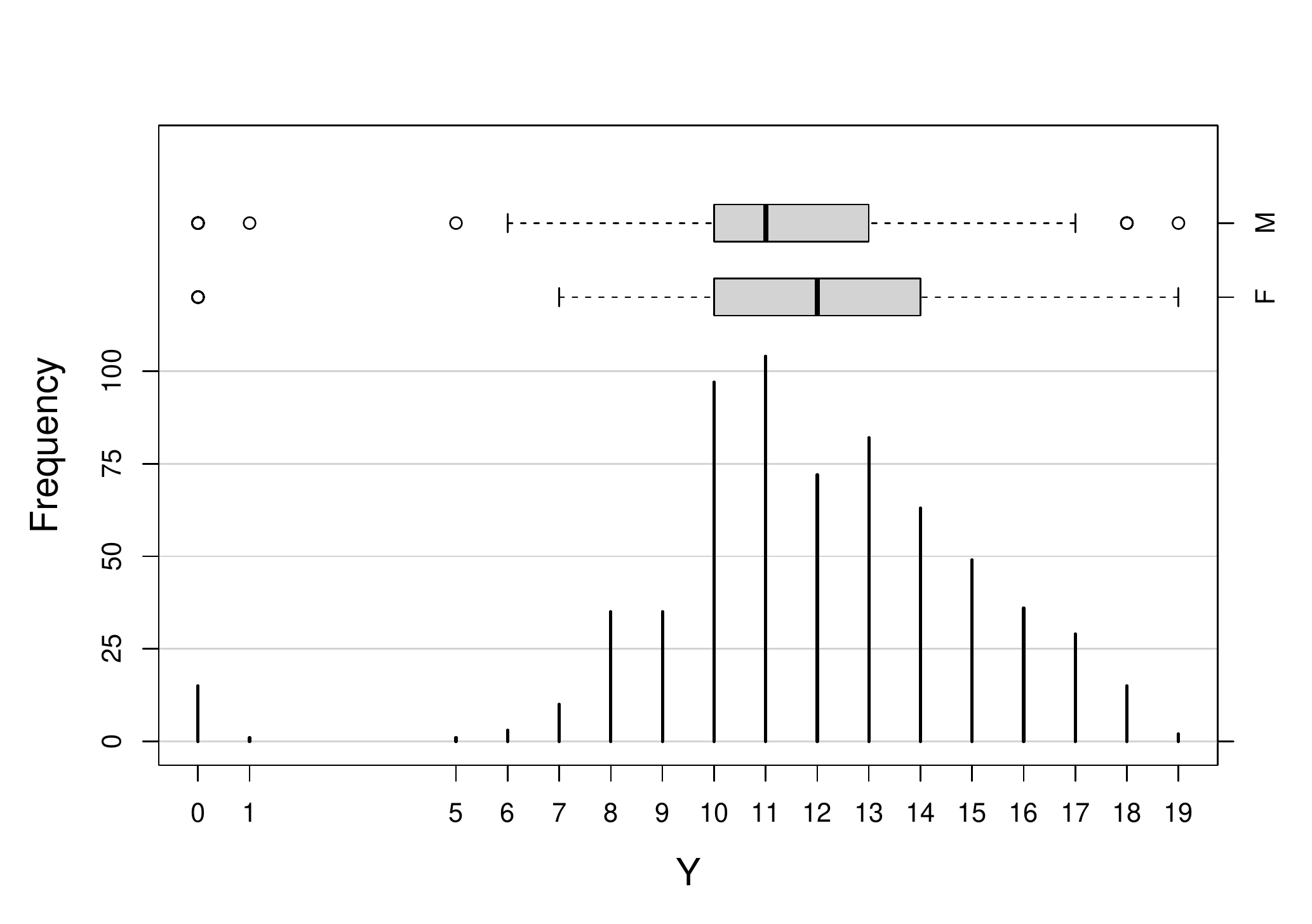}\\
  \caption{Frequency of final results (variable G3) in Portuguese language of $n=649$ students}\label{F1}
\end{figure}

As the dependent variable $y$ we consider the final grade with integer values ranging between 0 and 20 in Portuguese language (variable G3) of $n_{1} = 383$ female and $n_{2} = 266$ male students. The dummy variable $z$ takes values $0$ for female and $1$ for male. As also indicated by Figure \ref{F1}
female students perform better with an average of $\overline{y}_{1} = 12.25326$ compared to
$\overline{y}_{2} = 11.40602$ for male students. The corresponding equal variances two-sample test statistic admits $t=3.310938$ with p-value
$0.0009815287$. Although this implies strong significance the corresponding effect size from
(\ref{E2}) reads $d = 0.264261$, thereby indicating only a slightly more than low effect. This value may also
be obtained by function {\tt cohens\_d} from the {\sf R} package {\tt effectsize}, see \citet{R:effectsize}.

As additional independent variables we consider the education of the father (Fedu) $x_{1}$ and the travel time from home to school (traveltime) $x_{2}$. Both variables are measured on an ordinal scale with integer values ranging from 0 to 4 and 1 to 4, respectively, and are included as quantitative variables in our regression approach, implying $w=2$.

\begin{table}[hbt]
\centering
\begin{alltt}
                    Estimate Std. Error t value Pr(>|t|)
        (Intercept)  11.4138     0.4306  26.507  < 2e-16 ***
        ZM           -0.9406     0.2503  -3.759 0.000186 ***
        X1            0.6096     0.1144   5.329 1.37e-07 ***
        X2           -0.3369     0.1676  -2.010 0.044826 *
\end{alltt}
\caption{Extract from {\sf R} output by fitting the complete model (\ref{E4}) with function {\tt lm}, where variable sex is included as a factor, implying that {\tt ZM} represents the dummy $z$}\label{Tab1}
\end{table}
Least squares estimates of the coefficients with corresponding $t$ statistic values are given in Table \ref{Tab1}. The intercept estimate $\widehat{\beta}_{0}=11.41385$ is the average of the adjusted final grade $y_{\ast}$ of females, while the dummy variable estimate $\widehat{\beta}_{1}=-0.9406209$ is the difference between the average of $y_{\ast}$ in the male group minus the average in the female group as given in \ref{E18}. As it is also seen, better father's eduction comes along with better grades (positive $\widehat{\beta}_{2}$) while longer travel times to school come along with lower grades (negative $\widehat{\beta}_{3}$) when the other variables are held constant, respectively.  The coefficient of determination from this model reads $R^2 = 0.07238847$, while for the reduced model (omitting sex) it is $R_{0}^2 = 0.05207054$. Then $f^2$ from (\ref{E15}) is $f^2 = 0.0219035$, implying a slightly more than small effect concerning the difference between female and male final grades, ceteris paribus. % $x_{1}$ and $x_{2}$ are held constant.

The relationship (\ref{E14}) may also be used to infer Cohen's $d_{\ast}$ from $f^2$. For this
\begin{equation}
(X_{1}^{T} M_{2} X_{1})^{-1} = \begin{pmatrix}
0.019062813 & -0.001591796\\
-0.001591796 & 0.006438624
\end{pmatrix}\; ,
\end{equation}
the lower-right element being the scaled variance $\gamma$ of $\widehat{\beta}_{1}$. With $\widehat{\sigma} = 3.118756$ it follows $d_{\ast}= 0.3016013$.
This indicates a slightly stronger effect when variables $x_{1}$ and $x_{2}$ are held constant than seen before from  $d = 0.264261$ not considering any additional independent variables at all. Alternatively, $d_{\ast}$ may be computed by either of the formulas (\ref{E17}) or (\ref{E9}) yielding the very same absolute value.

\appendix

\section{Proofs of Theorems}

In this section we give short proofs of the three stated theorems. Although most of the supporting formulas and derivations may already be found throughout the literature, we present them here in order to provide a unified and self-contained treatment of the topic.

\begin{pot1}
From (\ref{E6}) it follows that
\begin{equation}
y - X_{1} \widehat{\delta}_{1} -  X_{2} \widehat{\delta} =(I_{n} - P) y\; ,
\end{equation}
where $P=X(X^{T}X)^{-1} X^{T}$ is the orthogonal projector onto the column space of the model matrix $X=(X_{1}, X_{2})$.
Since the column space of $X_{1}$ is contained in the column space of $X$, it follows
\begin{equation}
P P_{1} = P_{1} = P_{1} P
\end{equation}
for $P_{1} =X_{1} (X_{1}^{T} X_{1})^{-1} X_{1}^{T}$. Hence,
\begin{equation}
X_{1} \widehat{\delta}_{1} = P_{1} y = P_{1} (y - X_{2} \widehat{\delta}_{2})\; .
\end{equation}
Since $X_{1}$ is of full column rank, Theorem \ref{T1} follows.\qed
\end{pot1}

\begin{pot2}
From (\ref{E7})
\begin{equation}
y_{\ast} = y -X_{2}\widehat{\delta}_{2} = (I_{n} - X_{2} (X_{2}^{T} M_{1}  X_{2})^{-1} X_{2}^{T} M_{1}) y\; .
\end{equation}
Then
\begin{equation}\label{E13}
(y -X_{2}\widehat{\delta}_{2})^{T} M_{1} (y -X_{2}\widehat{\delta}_{2})
= y^{T} L y, \quad L= M_{1} - M_{1} X_{2}(X_{2}^{T} M_{1}  X_{2})^{-1} X_{2}^{T} M_{1}\; .
\end{equation}
%It is readily verified that $L$ is symmetric and idempotent.
From Theorem 1.3.2 in \citet{christensen2020plane} the expectation of the quadratic form
$y^{T} L y$ is given by
\begin{equation}
\text{E}(y^{T} L y) = \sigma^2 \text{trace}(L) + \mu^{T} A \mu
\end{equation}
with $\mu = X_{1} \delta_{1} +  X_{2} \delta_{2}$. From (\ref{E8}) it follows $\text{trace}(M_{1}) =n_{1} - 1 + n_{2} -1$ and therefore $\text{trace}(L) = n_{1} + n_{2} - 2- w$. In addition $L X_{1}=0$ and $L X_{2}=0$, implying
$\text{E}(y^{T} L y) = \sigma^2 (n_{1} + n_{2} - 2- w)$. This gives Theorem \ref{T2}.\qed
\end{pot2}

\begin{pot3}  Similarly to (\ref{E7}), the Frisch-Waugh-Lovell theorem states that
\begin{equation}\label{E16}
\widehat{\delta}_{1} = (X_{1}^{T} M_{2}  X_{1})^{-1} X_{1}^{T} M_{2} y,\quad M_{2} = I_{n}- X_{2} (X_{2}^{T} X_{2})^{-1} X_{2}^{T}\; .
\end{equation}
Hence, it is seen that $\widehat{\delta}_{1}$ follows a multivariate normal distribution with expectation vector $\delta_{1}$ and variance-covariance matrix $\sigma^2 (X_{1}^{T} M_{2}  X_{1})^{-1}$,  e.g. by Exercise 1.8 in \citet{christensen2020plane}.
Then $\widehat{\beta}_{1}$ follows a univariate normal distribution with expectation $\beta_{1}$ and variance $\sigma^2 \gamma$, were $\gamma$ is the lower-right element of the
$2\times 2$ matrix $(X_{1}^{T} M_{2}  X_{1})^{-1}$. If $\beta_{1} =0$, then \begin{equation} U = \widehat{\beta}_{1}/(\sigma \sqrt{\gamma})\end{equation}
follows a standard normal distribution. Let
\begin{equation}
V = y^{T} L y/\sigma^2
\end{equation}
with  $L$ from (\ref{E13}). It is readily verified that $L$ is symmetric and idempotent, so that  Theorem 1.3.3 in
\citet{christensen2020plane} implies that $V$ follows a central $\chi^2$ distribution with $\text{rank}(L) = \text{trace}(L) =n-2-w$ degrees of freedom. From
\begin{equation}
\widehat{\delta}_{1}  = F y, \quad F =(X_{1}^{T} X_{1})^{-1} X_{1}{T}(I_{n} - X_{2} (X_{2}^{T} M_{1}  X_{2})^{-1} X_{2}^{T} M_{1})
\end{equation}
it is seen that $F L=0$ implying that $Fy$ ad $y^{T} L y$ are independent, see Theorem 1.3.7 in \citet{christensen2020plane}. Then also  $U$ and $V$ are independent and
\begin{equation}
t_{\ast} = \frac{U}{\sqrt{V/(n-2-w)}} = \frac{\widehat{\beta}_{1}/\sqrt{\gamma}}{\widehat{\sigma}}
\end{equation}
follows a central $t$ distribution with $n-2-w$ degrees of freedom when $\beta_{1}=0$.\qed
\end{pot3}

\end{document}